\documentclass{WileyMSP-template}

\begin{document}


\title{Enhanced temperature sensing by multi-mode coupling in an on-chip microcavity system}

\maketitle


\author{Xueyi Wang$^\ddagger$}
\author{Tingge Yuan$^\ddagger$}
\author{Jiangwei Wu}
\author{Yuping Chen*}
\author{Xianfeng Chen}


\dedication{$^\ddagger$These authors contributed equally to this work.}

\begin{affiliations}
X. Wang, T. Yuan, J. Wu, Prof. Y. Chen, Prof. X. Chen\\
State Key Laboratory of Advanced Optical Communication Systems and Networks\\
School of Physics and Astronomy\\
Shanghai Jiao Tong University\\ 
Shanghai 200240, China\\
Email: ypchen@sjtu.edu.cn\\

Prof. X. Chen\\
Shanghai Research Center for Quantum Sciences\\
Shanghai 201315, China\\

Prof. X. Chen\\
Collaborative Innovation Center of Light Manipulations and Applications\\ Shandong Normal University\\
Jinan 250358, China
\end{affiliations}



\begin{abstract}

The microcavity is a promising sensor platform, any perturbation would disturb its linewidth, cause resonance shift or splitting. However, such sensing resolution is limited by the cavity's optical quality factor and mode volume. Here we propose and demonstrate in an on-chip integrated microcavity system that resolution of a self-referenced sensor could be enhanced with multi-mode coupling. In experiments, inter-mode  coupling strength is carefully optimized with a pulley waveguide and observed a resolution improvement of nearly $3$ times in frequency domain. While experiencing small refractive index change tuned by temperature, mode-coupled system shows a $7.2$ times sensitivity enhancement that is than the uncoupled system on the same chip and a very significant lineshape contrast ratio change as great reference for minor frequency shifts. This approach will help design microcavity sensors to improve detection sensitivity and resolution under limited manufacture precision.

\end{abstract}


\section{Introduction}
Optical microcavity as one of the building blocks of photonic integrated circuit has enabled a variety of applications including nonlinear optics[1-2], low-threshold laser[3-4] and single molecule detection[5-12], its small mode volume and high quality  factor ($Q$ factor). Especially in unlabeled sensing[13-19] or environmental monitoring[20-25] as a great supplement for medical and environmental research. On the other hand, decrease in the cavity's mode volume would increase radiation losses that is no longer neglectable[26] causing drop in its $Q$ factor. Overcoming such limitation by introducing new principles to microcavity systems thus become urgent. There have been works implementing microcavity lasers[27-28] to enhance light-matter interaction or by utilizing opto-mechnical coupling[29] that boosts sensing resolution by magnitudes.

Among all of the brand new solutions, one of the most approachable is by introducing mode coupling into the system[30], for that it causes little extra fabrication or experiment difficulties. When the two coupled modes are at weak-coupling region[31], their coherent interaction can optimize the spectrum's lineshape for efficient sensing[32]. Within one single cavity, the coupling condition can be satisfied by utilizing modes in different polarization of a micro-toroidal cavity[33] or by applying UV curable adhesive onto micro-bottle resonator to create a lossy mode[34], which achieved a 4.3 times refractive index change sensitivity amplification through its coupling with another discrete mode. Meanwhile micro-ring resonator is the ideal platform for on-chip integration, with as simple as a built-in Fabry–Pérot (F-P) cavity on its coupled waveguide[35], multi-mode coupling between cavities could be achieved. Such structure was first manufactured with polymer platform[36] that increased the sensitivity of solution refractive index by its sharp resonance slopes, and with a silicon-on-insulator chip as well[37] which realized a tunable lineshape fitting a variety of applications. Recently, mode coupling has also been controlled by scatterers to function at its exceptional point showing possibility for unprecedented sensitivity[38]. Thus mode coupling could be a handy improvement to already widely studied microcavity sensors.

In this work, we propose a design method to improve the resolution of microcavity sensors through multi-mode coupling with a compact, on-chip integrated micro-cavity system. Based on a waveguide to micro-racetrack structure supporting three resonance modes simultaneously and a pulley coupler with careful geometrical optimization, our design allows efficient and distinct inter-mode coupling at $1520$ nm to $1555$ nm band for both racetrack quasi-TE and TM modes leading to frequency shifts and sharp lineshape, which helps to distinguish two modes during self-referenced sensing and breaks the sensitivity's dependence on the $Q$ factor that microcavity sensors always suffer from. In frequency domain we achieved 3 times enhancement in resolution and a sensitivity of 44 $\rm pm^\circ C^{-1}$ that is 7.2 times higher
than the uncoupled structure on the same chip, with lineshape contrast ratio (LCR) of 24.1 times enhancement in the same time which function as great reference for minor turbulence. Our proposed approach will benefit applications in optical sensors that require integration and high sensitivity probing weak signals under a compromised fabrication efficiency.
\section{Theory}
Conventionally when two modes are weakly coupled, for instance one discrete mode and one continuous mode, the discrete mode would experience a frequency shift and linewidth sharpening determined by their detuned wavelength and coupling strength[31]. While the coupling includes two discrete modes simultaneously, it is very likely that they will experience different shifts for that they possess distinct coupling strength and eigenfrequencies. Thus by controlling the composition of three modes we could manipulate the relative frequency difference of them after coupling happened, that in certain scenarios would help us to distinguish two discrete modes with higher resolution. Here we first introduce the theory and how it compose with our on-chip system.

In the waveguide micro-ring resonator (WGMRR) three modes co-exist in the system as in \textbf{Figure 1a}, one waveguide (WG) mode reflected by built-in gratings, two micro-ring resonance (MRR) modes with quasi-TE and TM polarization. They possess different coupling efficiency $\kappa_j$, internal loss $\gamma_j$ and resonant frequency separately $\omega_j$ ($j=0$ for WG mode and $j=1,2$ for MRR quasi-TE, TM modes respectively as shown in \textbf{Figure 1a}). The system Hamiltonian should be,
\begin{equation}
     H_{SYS}=\sum_{j, j=0,1,2}\hbar\omega_ia_i^{\dag}a_i+i\hbar\kappa_1(a_0^{\dag}a_1-a_1^{\dag}a_0)+i\hbar\kappa_2(a_0^{\dag}a_2-a_2^{\dag}a_0)-ig\hbar(a_1^{\dag}a_2+a_2^{\dag}a_1).
\end{equation}
\begin{figure}[h]
\centering\includegraphics[width=10cm]{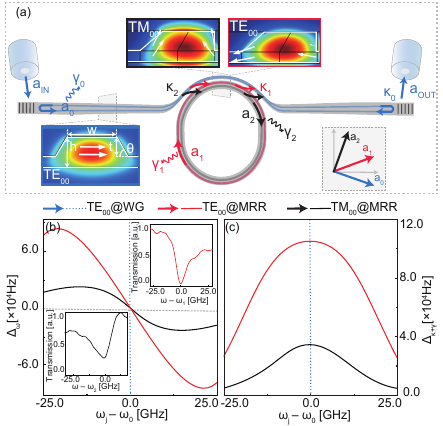}
\caption{a) Scheme of the WGMRR system with WG mode in blue and MRR quasi-TE and quasi-TM in red and black respectively. The insets are calculated with conformal transformation[39] indicating the polarization of $\rm TE_{00}$ and $\rm TM_{00}$ in straight and curved waveguides (see also Supporting Information section 1.1). Frquency shift b) and linewidth sharpening c)  for different MRR modes caused by coupling, insets are transmission spectrum of quasi-TE, TM modes after coupling.}
\end{figure}

In which $a_j^{\dag}$ $(a_j)$ are photon creation (annihilation) operators. For the oblique nature of the side walls ridge waveguide and birefringence of X-cut LN (Lithium Niobate), the quasi-TE (TM) modes are not perfectly parallel(vertical) to the substrate plane (see inset of \textbf{Figure 1a} and Supporting Information section 1.1) which causes them to couple with a coefficient $g$[40]. At the same time enabling WG $\rm TE_{00}$ photon $a_0$ to generate MRR TM mode $a_2$ with different polarization. Similarly, at the time $a_2$ couples back into the waveguide it is projected into TE polarized beams that lead to its coherent interaction with $a_0$ lights. 

Under the first Markov approximation $\kappa_0^2(\omega)=\kappa_0/2\pi$[41], we approach the Langevin equations of motion as follow,
\begin{equation}
   \left( \begin{array}{c}
         \dot{a_0}\\\dot{a_1}\\\dot{a_2}\end{array}\right)=
         \left(\begin{array}{ccc}
            -i\omega_0-\frac{\kappa_0+\gamma_0}{2} & \kappa_1&\kappa_2  \\-\kappa_1& -i\omega_1-\frac{\kappa_1+\gamma_1}{2} & -g\\
             -\kappa_2 & -g& -i\omega_2-\frac{\kappa_2+\gamma_2}{2}\end{array}\right)\left( \begin{array}{c}
         a_0\\a_1\\a_2\end{array}\right)+\left( \begin{array}{c}
         \sqrt{\kappa_0}a_{IN}\\0\\0\end{array}\right),
\end{equation}
$a_{IN}=\sqrt{2P_{IN}\kappa_0/\hbar\omega_0}$ is the input amplitude in TE polarization and so as $a_0$. Then we solve Equation (2) in frequency domain,
\begin{equation}
    (\omega-\widetilde{\omega}_0)a_0=-i\kappa_1a_1-i\kappa_2a_2-i\sqrt{\kappa_0}a_{IN}
\end{equation}
\begin{equation}
    (\omega-\widetilde{\omega}_1)a_1=iga_2+i\kappa_1a_0,
\end{equation}
\begin{equation}
    (\omega-\widetilde{\omega}_2)a_2=iga_1+i\kappa_2a_0,
\end{equation}
in which $\widetilde{\omega}_j=\omega_j-i\frac{\kappa_j+\gamma_j}{2}(j=0,1,2)$ is the complex eigenfrequency. Equation (3) implies that the output field is made up by the coherent superposition of $a_1,a_2$ to $a_0$ mode with coupling constant $\kappa_1$ and $\kappa_2$ respectively.  For that inter-modal coupling coefficient $g$ within the MRR is  relatively small, output amplitude $a_{OUT}=\sqrt{\kappa_0}a_0-a_{IN}$ approximates to,
\begin{equation}
   a_{OUT}=\kappa_0\xi a_{IN}-a_{IN},
\end{equation}
\begin{equation}
    \xi=\frac{i(\omega-\widetilde{\omega}_1)(\omega-\widetilde{\omega}_2)}{(\omega-\widetilde{\omega}_0)(\omega-\widetilde{\omega}_1)(\omega-\widetilde{\omega}_2)-\kappa_1^2(\omega-\widetilde{\omega}_2)-\kappa_2^2(\omega-\widetilde{\omega}_1)},
\end{equation}
the transmission spectrum should be,
\begin{equation}
    T=\left|\frac{a_{OUT}}{a_{IN}}\right|^2\approx1-2\kappa_0\xi,
\end{equation}
where $|\kappa_0\xi|^2$ is ignored in above equation. Solving the denominator of $\xi$ there are (details in Supporting Information 2), 
\begin{equation}
        \widetilde{\omega}_{j\pm}=(\widetilde{\omega}_0+\widetilde{\omega}_j\pm\delta_j)/2, \delta_j^2=(\widetilde{\omega}_0-\widetilde{\omega}_j)^2+4\kappa_j^2 ,
\end{equation}
the eigenfrequencies of MRR are shifted by complex frequencies $\widetilde{\Delta}_{\pm j}=\widetilde{\omega}_{j\pm }-\widetilde{\omega}_j=(\widetilde{\omega}_0-\widetilde{\omega}_1\pm\delta_j)/2$, $(+:\omega_0-\omega_j>0, -:\omega_0-\omega_j<0)$ respectively, in which $\Delta_{\omega}=Re(\widetilde{\Delta}_{\pm j})$ stands for shifts in frequency and $\Delta_{\kappa+\gamma}=-Im(\widetilde{\Delta}_{\pm j})$ for changes in linewidth. Consequently, MRR modes experience a red (blue) shift if they are blue (red) detuned to $\omega_0$ and a linewidth reduction either way as shown in \textbf{Figure 1b} and \textbf{c}. For that quasi-TE mode has greater $\kappa$ that leads to bigger $|\Delta_{\omega}|$ and $|\Delta_{\kappa+\gamma}|$ for $a_1$. 

Consequently, if two MRR modes locate across the continuum mode eigenfrequency they would be "pulled closer" as their frequencies are shifted towards each other $|\Delta'_{12}|=|\Delta_{12}|-\left[|Re(\delta_1)|+|Re(\delta_2)|\right]/2$ as case \textbf{I} in \textbf{Figure 2a}. Or they could be "pushed apart" if coupled to  two different continuum modes as the cases in \textbf{Figure 2a II}. that $|\Delta'_{12}|=|\Delta_{12}|+\left[|Re(\delta_1)|+|Re(\delta_2)|\right]/2$. In the case when their frequency differs even less, within one side of the background's FSR (free spectrum range), then they are "pushed apart" when mode with larger $\kappa$ locates closer to $\omega_0$ only then $|\Delta'_{12}|=|\Delta_{12}|+\left[|Re(\delta_1)|-|Re(\delta_2)|\right]/2$ could be enlarged (as the case in \textbf{Figure 2a III}, the width of modes' stripes indicates their relative coupling strength), and "pulled closer" if in the opposite case. Under the circumstances when multi-mode coupling leads to "pushed apart", it enhances the observation resolution of two adjacent MRR modes, which is ideal for sensing applications or mode measurements. From our following experiments, clearly in \textbf{Figure 2b}, \textbf{c} and \textbf{d} the transmission spectrum of the WGMRR and WG (in black and blue) has strongly asymmetric dips and peaks indicate multi-mode coupling happening in the above (\textbf{I} to \textbf{III}) scenarios.
\begin{figure}[htb]
\centering\includegraphics[width=\linewidth]{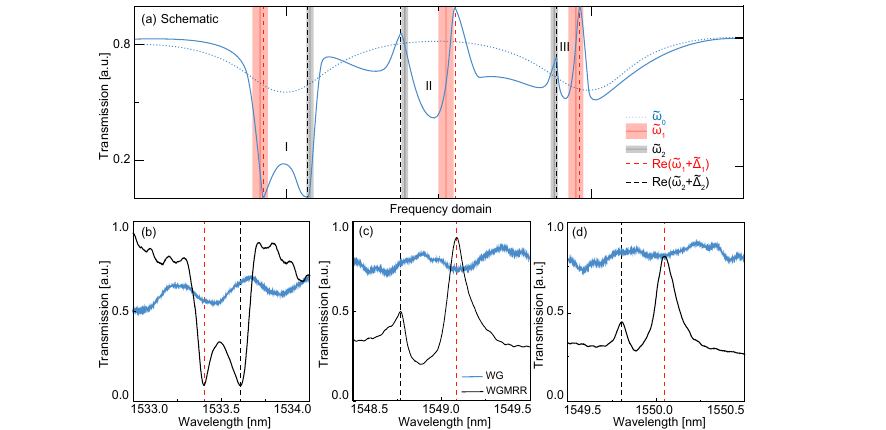}
\caption{a) Schematic of the proposed mode coupling involving three modes arranged in 3 fashions leading to different mode resolution: when MRR modes locate \textbf{I} across background eigenmode then they are "pulled closer", \textbf{II} over different background eignmodes and \textbf{III} within one side of background mode while mode with larger $\kappa$ is closer to $\omega_0$ when they are "pushed apart", wider stripe indicates stronger coupling strength of the eigenmode to the background mode. Transmission spectrum of the WGMRR and WG in black and blue which satisfies the above scheme \textbf{I} b), \textbf{II} c), and \textbf{III} d) respectively.}
\end{figure}
\section{Results}
\subsection{Exprimental setup}
\begin{figure}[htb]
\centering\includegraphics[width=\linewidth]{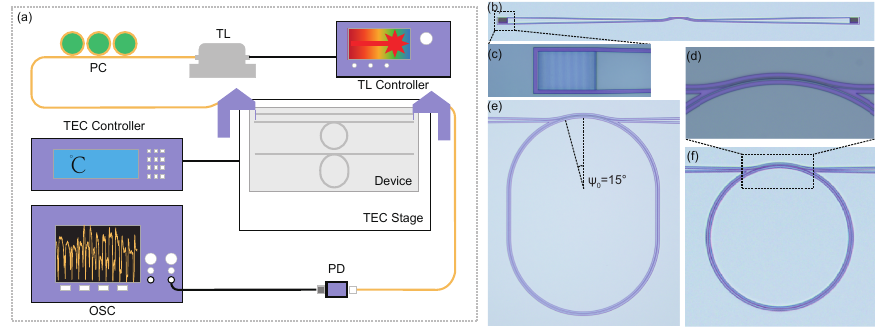}
\caption{a) Experiment setup. TL: Tunable Laser, PC: polarization controller, PD: photon detector, OSC: oscillator, TEC: thermoelectric cooler, electrical wire and optical fiber in black and yellow respectively. Microscopic pictures of the single waveguide b), gratings c), coupling region d), racetrack e) and ring f) resonators. }
\end{figure}
The experimental setup is in \textbf{Figure 3a}, it involves a tunable infrared laser source (New Focus TLB-6728) which is adjusted by a PC (polarization controller) and then coupled into the chip through built-in gratings on the waveguide. After the output port a PD (photodetector) collects the transmission information that is then shown on the OSC (oscilloscope). Meanwhile the chip is loaded on a TEC (thermoelectric cooler) stage with a precision up to $0.01$ degree Celsius. From this setup any multi-mode coupling effect is acquired from the the lineshape of transmission spectrum.

To achieve multi-mode coupling and compare its effects, we integrate a waveguide, a waveguide to micro-ring and a waveguide to micro-racetrack all manufactured on one single X-cut LNOI (lithium niobate on insulator) chip with standard electron beam lithography and plasma-reactive etching (see fabrication details in Supporting Information section 3). As marked in insets of \textbf{Figure 1a}, the chip thickness is $h=0.6$ $\mu m$, while top width of waveguide is $w=1$ $\mu m$, thickness $t=0.38$ $\mu m$ and side wall angle $\theta=60^{\circ}$, which are ideal to support only fundamental modes. The radius of both micro-racetrack and micro-ring is the same $R=129.03$ $\mu m$ while the racetrack contains an extra straight waveguide of $82.54$  $\mu m$. At the coupling region \textbf{Figure 3c}, the waveguide width shrinks down to $0.8$ $\mu m$ with a gap of $G=0.6$ $\rm\mu m$ to achieve sufficient evanescent field coupling, next we calculate such coupling strength and analyse its effect in detail.
\subsection{Multi-mode coupling in frequency domain}
From the system Hamiltonian we could tell that the inter-modal coupling strength is determined by the mode coupling efficient $\kappa$ at the pulley waveguide, and it can be calculated with temporal perturbation theory as[42] (details in Supporting Information section 1.2),
\begin{equation}
    \kappa=\int_{-\psi_0}^{\psi_0}\left[\frac{i\omega}{4}\int_0^{R+G}\int_0^t\left(\epsilon-\epsilon_0\right)\mathbf{E}_{WG}\cdot\mathbf{E}_{MRR} rdrdz\right]e^{i\varphi} d\psi,
\end{equation}
in which $\mathbf{E}_{WG}\cdot\mathbf{E}_{MRR}$ corresponds to the mode overlap of normalized WG and MRR fields, permittivity $\epsilon$ can be obtained from the electric field and is shown in the mode's effective refractive index ($n_{eff}$) in \textbf{Figure 4a } and \textbf{b}. And $\varphi=k_0n_{WG}R_{WG}\psi-m\psi$, $m$ is the radial mode number of resonance mode inside cavity, $\varphi$ reflects the phase mismatch between the waveguide and cavity mode that has major impact on $\kappa$. Meanwhile X-cut LN is a anisotropic material as $n_{eff}$ shifts across azimuth angle $\psi$ infecting phase matching condition along its way, thus it introduces change in $\kappa$ proportional to $sinc(\varphi)$, see \textbf{Figure 4c}, where $\kappa_{TE}$ increases and $\kappa_{TM}$ decrease as $\psi$ grows. It offers another degree of freedom to manipulate coupling strength between different polarization with the angle and length of pulley coupling scheme. Also in frequency domain (\textbf{Figure 4d}) $\kappa$ degenerates significantly at longer input wavelength due to non-ideal phase matching and mode overlaping, in the following experiments we clearly observed weaker mode coupling at longer wavelength as shown in \textbf{Figure 4g}. Reminding that the pulley waveguide needs to be carefully designed to achieve sufficient coupling at target work waveband. In our design, $\kappa$ is set to generate sufficient coupling for both polarization across C-band at the same time differs with nearly one degree of magnitude so they appear with diverse mode shifts during coupling to be distinguished.

\begin{figure}[!h]
\centering\includegraphics[width=\linewidth]{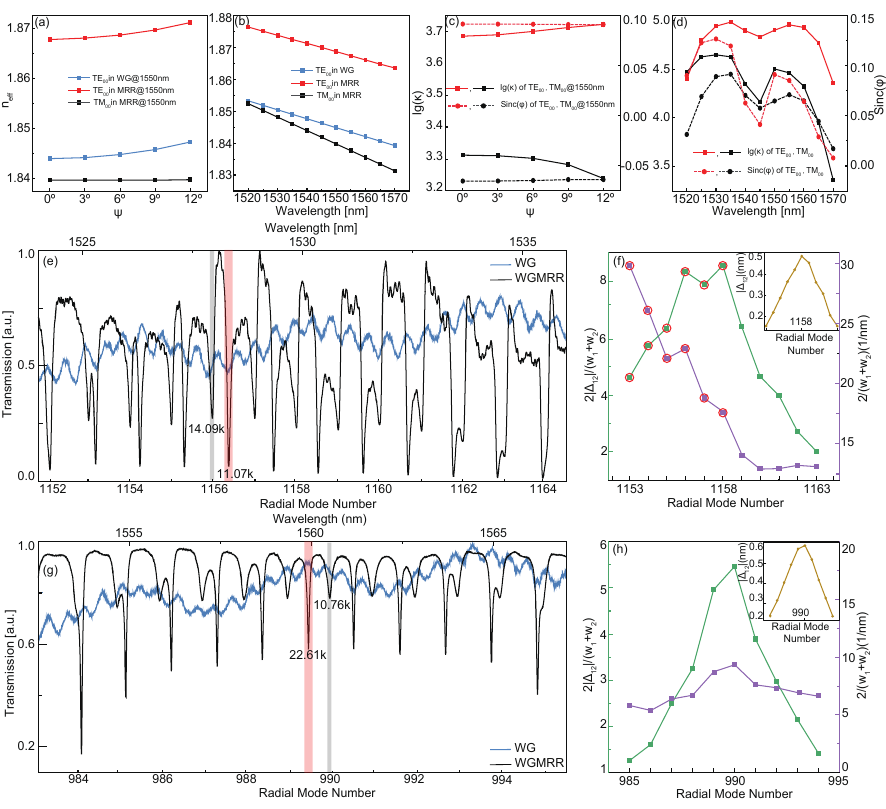}
\caption{The simulated effective refractive index ($n_{eff}$) at different angles a) on the X-cut LN chip and across resonance wavelength b). Coupling coefficient $lg(\kappa)$ and $sinc(\varphi)$ at different angles c) on the X-cut LN chip and across resonance wavelength d). e)Transmission spectrum of the WG-micro-racetrack and the sole waveguide (in blue) when there is multi-mode coupling. g)Transmission spectrum of the WG-micro-ring and the sole waveguide (in blue) when there is  no multi-mode coupling. Measured wavelength difference $|\Delta_{12}|$, FWHM $2/(w_1+w_2)$ and their product $2|\Delta_{12}|/(w_1+w_2)$ declaring the degree of mode separation of TM mode and its closest TE mode in frequency domain for coupled f) and uncoupled h) system, the red circle marks where inter-modal separation is significantly enlarged by coupling.}
\end{figure}
With above system we first obtain the transmission spectrum of the micro-racetrack (\textbf{Figure 4e}) the cavity modes preserved a strongly asymmetric lineshape due to the coupling with WG modes (the transmission spectrum of waveguide is in blue). The effective $Q$ factor calculated by applying Lorentz fitting of the dip reaches $Q_E=11.07k$ and $Q_M=14.09k$ for TE and TM respectively. With calculated coupling $Q_{CE}=17.10k$, $Q_{CM}=62.31k$ which ends up in coupling coefficients $\kappa_E=2.67\times10^5$, $\kappa_M=1.40\times10^5$ slightly larger than the simulation in \textbf{Figure 4d} supposedly caused by coupling region $\psi>15^{\circ}$ which has a larger gap but still allows evanescent field coupling. Compared to their intrinsic $Q$ factors $Q_{0E}=27.52k$, $Q_{0M}=18.21k$, the TE modes are over-coupled leading to strong interference by WG modes, theoretically it causes mode shift up to $72.52$ kHz as shown in \textbf{Figure 1b}. Experimentally, due to the different FSRs of TE and TM modes, the wavelength difference between two closest modes $|\Delta_{12}|$ would first increase and then decrease to zero during a certain wavelength range as the insets of both \textbf{Figure 4f} and \textbf{h}. In a coupled system, $|\Delta_{12}|$ and modes' FWHM  (full width half maximum, $w_1$, $w_2$ for TE and TM respectively) were tuned by coupling strength and relative background phase as in \textbf{Figure 4f}, determined by $\widetilde{\omega}_0-\widetilde{\omega}_1$ according Equation (9). The measured degree of mode separation $\frac{|\Delta_{12}|}{(w_1+w_2)/2}$ is enhanced for TM modes with radial mode number $1153$ to $1158$ (marked in red in \textbf{Figure 4f}), satisfying either scenario \textbf{II} or \textbf{III} from \textbf{Figure 2a}, nearly 3 times larger compared to their siblings (mode number $1153$ to $1163$).

Meanwhile in the uncoupled system with micro-ring resonator (\textbf{Figure 3e}), whose loaded $Q$ factors are $Q_{E}'=22.61k$, $Q_{M}'=10.76k$ and coupling $Q_{CE}'=137.37k$, $Q_{CM}'=203.82k$ that is significantly under-coupled. Calculated from its spectrum \textbf{Figure 4g} the mode separation $\frac{|\Delta_{12}|}{(w_1+w_2)/2}$ is proportional to $|\Delta_{12}|$ despite of the fluctuation in FWHM caused by coupling depth as in \textbf{Figure 4h} showing no signs of resolution enhancement nor inter-mode coupling. Next we run tests over their sufficiency of detecting minor temperature shifts for both systems.

\subsection{Temperature sensing enhanced by multi-mode coupling}
\begin{figure}[htb]
\centering\includegraphics[width=\linewidth]{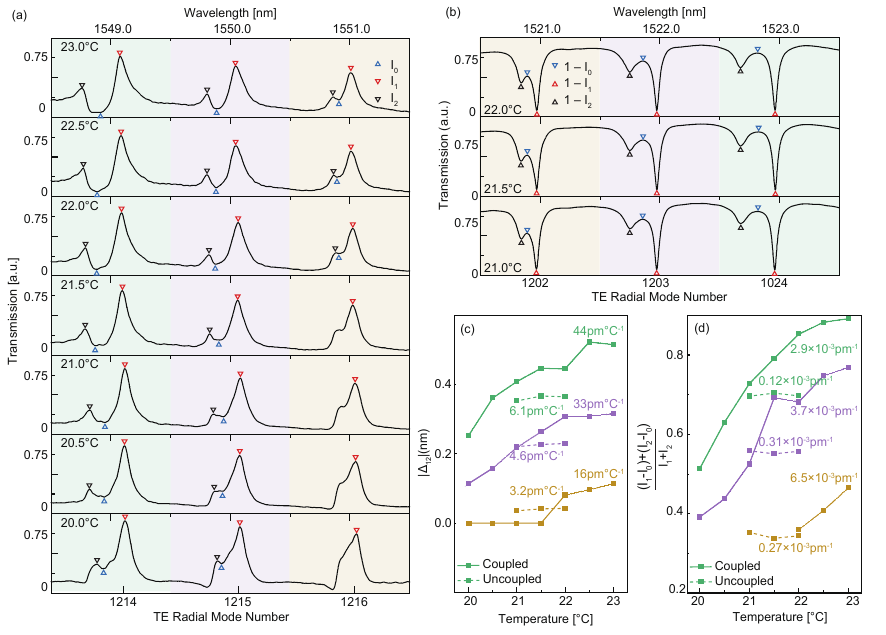}
\caption{a) Transmission spectrum the WGMRR with multi-mode coupling leading to "pushing apart". b)  Transmission spectrum the WGMRR without mode coupling by adjusting input polarization and wavelength. c) Measured wavelength separation $|\Delta_{12}|$ of a) and b)(dashed) at different temperature. d) Calculated lineshape contrast ratio $\frac{(I_1-I_0)+(I_2-I_0)}{I_1+I_2}$ of a) and b)(dashed), under multi-mode coupling their lineshapes experience a contrast shift $24.1$ times greater.  }
\end{figure}
In previous works utilizing double mode coupling such as fano resonance for sensing purpose, apart from the sharp asymmetric lineshape that declines the effective mode linewidth, those two modes are self-referenced so the system accuracy is no longer limited by experiment equipment[42]. According to above analysis, MRR quasi-TE and TM modes would sufficiently form a self-referenced sensing system while coupling with WG mode further improves its capacity. Here to compare the sensitivity for weak turbulence of a mode-coupled and an uncoupled system, we keep the laser source sweeping in wavelength and slightly adjust the stage's temperature control which tunes $n_{eff}$ of the MRR consequently. At around $1550$ nm wavelength the micro-racetrack forms an EIT-like spectrum in \textbf{Figure 5a}, the three groups of modes experienced different shifts due to their coupled background phase, the gap between two closest modes $|\Delta_{12}|$ shifts as large as $44$ $\rm pm^\circ C^{-1}$ which is $7.2$ times larger than the uncoupled system in \textbf{Figure 5b} that has very similar mode gaps in the first place, as shown in \textbf{Figure 5c}. 

On the other hand, a mode-coupled system possess a lineshape sensitive to its background phase[43]. In our setup the background modes have a linewidth of $0.23$ nm and FSR of $0.45$ nm (and it is also insensitive to thermal changes, see Supporting Information section 4), thus even a minor shift at $\sim \rm pm$ would lead to observable changes to the spectrum lineshape. That experiences a change in its LCR $\left(\frac{(I_1-I_0)+(I_2-I_0)}{I_1+I_2}\right)$ as large as $6.46\times10^{-3}$ $\rm pm^{-1}$, while the according uncoupled modes only changes $0.27\times10^{-3}$ $\rm pm^{-1}$ as in \textbf{Figure 5d}. So there is a maximum $24.1$ times enhancement compared with the uncoupled spectrum at similar temperature working as another crucial criterion for minor turbulence in the system's $n_{eff}$.

It is still worth noting that our design did not reach the best performance among all of the microcavity thermal sensors. Apart from the fact that our on-chip microcavities were built large in the first place to compensate the WG mode's FSR, the pulley waveguide was designed to reach over-coupling condition which lead to decrease in MRR modes' $Q$ factor and coupling depth.  Taking those information under consideration, a further improvement should utilize the coupling of multiple high-$Q$ factor modes either by connecting several micro-resonators[30] or by optimizing cavities that support a number of resonance modes.

\section{Conclusion}

In this paper, we have in depth analysed dimensions of multi-mode coupling, revealing the intrinsic connection between frequency, $Q$ factor enhancement and modes' composition. Based on theoretical discussions and experiments demonstrated with an on-chip integrated WGMRR, we have achieved higher resolution in frequency domain and improved sensitivity as a self-referenced sensor with micro-racetrack's quasi-TE and TM modes coupled to waveguide resonance mode at the same time. Here we have confined our discussions of mode-coupled sensors within the fundamental case, practiced between discrete and continuum modes, and it is expected that a natural extension to more sophisticated configurations with several high-$Q$ modes or even multiple exceptional points would reach much higher magnitudes of enhancement for microcavity sensors without over-investing into manufacture techniques. Though our design was practiced with a LNOI micro-racetrack, we believe that this method can be applied to any cavity-based sensors or other material platforms with great integration capability.

\medskip
\textbf{Supporting Information} \par 
Supporting Information is available from the Wiley Online Library or from the author.

\medskip
\textbf{Acknowledgements} \par 
The authors would like to acknowledge support from National Natural Science Foundation of China (Grant Nos. 12134009) and SJTU (No. 21X010200828). 

\medskip
\textbf{Conflict of Interest} \par 
The authors declare no conflict of interest. 

\medskip
\textbf{Data Availability Statement} \par 
The data that support the findings of this study are available from the cor-responding author upon reasonable request.

\medskip
\textbf{Keywords} \par 
Micro-optical device, Optical sensing and sensors, Mode coupling, Integrated optics, Resonant modes

\medskip
%

\textbf{References}\\

[1] T. J. Kippenberg, S. M. Spillane, K. J. Vahala, Phys. Rev. Lett. 2004, 93 083904\\

[2] X. Ye, S. Liu, Y. Chen, Y. Zheng, X. Chen, Opt. Lett. 2020, 45, 2 523.\\

[3] J. W. B. Z. Y. C. . X. C. YiAn Liu, XiongShuo Yan, Sci. China Phys. Mech. Astron. 2021, 64, 64234262.\\

[4] X. Liu, X. Yan, Y. Liu, H. Li, Y. Chen, X. Chen, Opt. Lett. 2021, 46, 21 5505.\\

[5] F. Vollmer, L. Yang, Nanophotonics 2012, 1, 3-4 267.
[6] Y. Zhi, X.-C. Yu, Q. Gong, L. Yang, Y.-F. Xiao, Advanced Materials 2017, 29, 12 1604920.\\

[7] X. Jiang, A. Qavi, S. Huang, L. Yang, Matter 2020, 3, 2 371.\\

[8] M. R. Foreman, J. D. Swaim, F. Vollmer, Adv. Opt. Photon. 2015, 7, 2 168.\\

[9] S.-J. Tang, M. Zhang, J. Sun, J.-W. Meng, X. Xiong, Q. Gong, D. Jin, Q.-F. Yang, Y.-F. Xiao, Nature Photonics 2023, 1–6.\\

[10] F. Vollmer, S. Arnold, D. Keng, Proceedings of the National Academy of Sciences 2008, 105, 5220701.\\

[11] X. Y.-F. L. L. H. L. C. D.-R. Zhu Jiangang, Ozdemir Sahin Kaya, Y. Lan, Nature Photonics 2010, 4, 1 46.\\

[12] T. Lu, H. Lee, T. Chen, S. Herchak, J.-H. Kim, S. E. Fraser, R. C. Flagan, K. Vahala, Proceedings of the National Academy of Sciences 2011, 108, 15 5976.\\

[13] S. Frustaci, F. Vollmer, Current Opinion in Chemical Biology 2019, 51 66, chemical Genetics and Epigenetics • Molecular Imaging.\\

[14] F. Vollmer, S. Arnold, D. Keng, Proceedings of the National Academy of Sciences 2008, 105, 5220701.\\

[15] F. Vollmer, S. Arnold, Nature Methods 2008, 5, 7 591.\\

[16] R. W. Boyd, J. E. Heebner, Appl. Opt. 2001, 40, 31 5742.\\

[17] F. Vollmer, D. Braun, A. Libchaber, M. Khoshsima, I. Teraoka, S. Arnold, Applied Physics Letters 2002, 80, 21 4057.\\

[18] W. Kim, S. K. Ozdemir, J. Zhu, F. Monifi, C. Coban, L. Yang, Opt. Express 2012, 20, 28 29426.\\

[19] O. Gaathon, J. Culic-Viskota, M. Mihnev, I. Teraoka, S. Arnold, Applied Physics Letters 2006, 89, 22 223901.\\

[20] A. M. Armani, K. J. Vahala, Opt. Lett. 2006, 31, 12 1896.\\

[21] W. Kim,K.$\rm\ddot{O}$zdemir, J. Zhu, L. He, L. Yang, Applied Physics Letters 2010, 97, 7 071111.\\

[22] Q. Lu, X. Chen, L. Fu, S. Xie, X. Wu, Nanomaterials 2019, 9, 3.\\

[23] X. Xu, W. Chen, G. Zhao, Y. Li, C. Lu, L. Yang, Light: Science and Applications 2018, 7, 162.\\

[24] C.-H. Dong, L. He, Y.-F. Xiao, V. R. Gaddam, S. K. Ozdemir, Z.-F. Han, G.-C. Guo, L. Yang, Applied Physics Letters 2009, 94, 23 231119.\\

[25] J. Liao, L. Yang, Light: Science and Applications 2021, 10, 1 32.\\

[26] C. Ciminelli, F. Dell’Olio, D. Conteduca, C. Campanella, M. Armenise, Optics Laser Technology 2014, 59 60.\\

[27] L. He, K.$\rm\ddot{O}$zdemir, J. Zhu, W. Kim, L. Yang, Nature Nanotechnology 2011, 6, 7 428.\\

[28] K.$\rm\ddot{O}$zdemir, J. Zhu, X. Yang, B. Peng, H. Yilmaz, L. He, F. Monifi, S. H. Huang, G. L. Long, L. Yang, Proceedings of the National Academy of Sciences 2014, 111, 37 E3836.\\

[29] W. Yu, W. C. Jiang, Q. Lin, T. Lu, Nature Communications 2016, 7, 1 12311.\\

[30] Y.-F. Xiao, V. Gaddam, L. Yang, Opt. Express 2008, 16, 17 12538.\\

[31] B. Peng, K.$\rm\ddot{O}$zdemir, W. Chen, F. Nori, L. Yang, Nature Communications 2014, 5, 1 5082.\\

[32] C.-M. Chang, O. Solgaard, Opt. Express 2013, 21, 22 27209.\\

[33] B.-B. Li, Y.-F. Xiao, C.-L. Zou, Y.-C. Liu, X.-F. Jiang, Y.-L. Chen, Y. Li, Q. Gong, Applied Physics Letters 2011, 98, 2, 021116.\\

[34] J. Liao, X. Wu, L. Liu, L. Xu, Opt. Express 2016, 24, 8 8574.\\

[35] S. Fan, Applied Physics Letters 2002, 80, 6 908.\\

[36] C.-Y. Chao, L. J. Guo, Applied Physics Letters 2003, 83, 8 1527.\\

[37] L. Gu, H. Fang, J. Li, L. Fang, S. J. Chua, J. Zhao, X. Gan, Nanophotonics 2019, 8, 5 841.\\

[38] W. Chen, K.$\rm\ddot{O}$zdemir, G. Zhao, J. Wiersig, L. Yang, Nature 2017, 548, 7666 192.\\

[39] M. Heiblum, J. Harris, IEEE Journal of Quantum Electronics 1975, 11, 2 75.\\

[40] L. Cortes-Herrera, X. He, J. Cardenas, G. P. Agrawal, Phys. Rev. A 2021, 103 063517.\\

[41] C. Gardiner, P. Zoller, Quantum noise: a handbook of Markovian and non-Markovian quantum stochastic methods with applications to quantum optics, Springer Science and Business Media, 2004.\\

[42] S.-L. Chuang, Journal of Lightwave Technology 1987, 5, 1 5.\\

[43] U. Fano, Phys. Rev. 1961, 124 1866.\\

\end{document}